# Diurnal Variations of Lee Wave Clouds on Mars using Emirates eXploration Imager (EXI)

Mariam R. Alshamsi[1,2] , Mashhoor A. Al-Wardat [2,3]

[1] Mohammed Bin Rashid Space Centre (MBRSC), P.O. Box 211833, Dubai, UAE

[2] Department of Applied Physics and Astronomy, College of Sciences, University of Sharjah, POBox 27272, Sharjah, UAE

[3] Sharjah Academy for Astronomy, Space Sciences and Technology, University of Sharjah, POBox 27272, Sharjah, UAE

u20105442@sharjah.ac.ae, malwardat@sharjah.ac.ae,

## Abstract

Understanding the diurnal behavior of lee wave clouds on Mars provides critical insight into the planet's mesoscale atmospheric dynamics and their interaction with surface topography. Lee wave clouds exhibit distinct spatial and temporal patterns that vary over the Martian day. In this study, we investigate the diurnal distribution and frequency of lee wave cloud activity during Martian Year (MY) 36 using observations from the EXI instrument aboard the Emirates Mars Mission (EMM) "Hope" spacecraft. A total of 50 lee wave events were identified, with a pronounced peak in activity during afternoon hours between solar longitudes (Ls) 270°–360°. Our analysis reveals a seasonal and local-time dependence for these clouds, providing a comparative framework with previous mission datasets. These findings not only enhance the current understanding of Martian weather processes but also support future efforts to model and predict terrain-induced cloud dynamics across key locations on Mars.

# 1. Introduction

Studying the diurnal behavior of lee wave clouds on Mars offers insights into aspects of the dynamics of the atmosphere. Lee wave clouds form as the air mass flows over elevated terrain and exhibit unique patterns that evolve during the Martian day. Studies of lee wave clouds connect daily weather patterns with the interactions of surface topography and atmospheric processes. As a result, studying these features is expected to advance knowledge and improve the impact of current and future missions.

## 1.1 Mars atmosphere

Mars' atmospheric dynamics are a fascinating interaction of numerous factors that influence the behavior and evolution of its thin envelope. The low atmospheric pressure and low temperatures result in relatively weak winds compared to those on Earth, yet the planet experiences powerful dust storms that can engulf its entire surface. Seasonal variations also influence atmospheric dynamics, with carbon dioxide ice at the poles expanding and contracting, driving atmospheric circulation patterns. Furthermore, Mars' topography and surface features have a significant impact on its atmospheric behavior, impacting wind patterns and the production of phenomena such as lee wave clouds. [1].

## 1.2 Lee Wave clouds

Martian lee wave clouds are formations that arise due to the interaction between the Martian atmosphere and the planet's surface topography. As a stable air mass flows over elevated features, such as mountains or crater rims, it generates atmospheric structures in the form of wave-like oscillations [2]. When the air rises over these terrains, it cools adiabatically, resulting in the condensation of water vapor and the formation of these clouds. This phenomenon is generated by topographic features, such as crater rims or mountains, which have been observed to vary in height from as low as 10 km [3] to as high as 45 km [4].

## 1.3 Properties

Lee wave clouds on Mars exhibit distinct properties that vary by season, and potentially by time of day. A recent study has demonstrated a strong correlation between topographic altitude and the vertical distribution of zonal wind [3]. With strong static stability near the ground and an increase in wind speed with height through the troposphere, this supports the trapping of internal gravity waves within low levels of the atmosphere, forming lee wave clouds [5]. Lee wave clouds on Mars are mainly observed downwind of elevated topographical features, such as mountains and crater rims, where the interaction of the prevailing winds with the terrain generates the atmospheric waves necessary for their formation.

## 2.  Previous Missions

Studying Martian lee wave clouds has progressed our understanding of the planet's atmospheric dynamics through various planetary missions. Starting in 1971, Mariner 9 successfully became the first spacecraft to enter orbit around Mars, providing groundbreaking imagery of the planet's surface and atmosphere. It captured early evidence of orographic cloud formations near the Tharsis Montes, including wave-like patterns consistent with lee wave activity [6]. Following this, the Viking missions launched in 1975 significantly advanced atmospheric studies on Mars, with their instruments capturing layered cloud formations and wave-like structures in the vicinity of major volcanic terrains [7]. In 1997, the Mars Global Surveyor (MGS) began operations with its Mars Orbiter Camera (MOC), significantly improving spatial and temporal resolution. MGS captured seasonal cloud formations over elevated terrain and helped establish mesoscale modeling efforts for wave-cloud generation [8]. The European Space Agency's (ESA) Mars Express, launched in 2003, contributed high-resolution stereo imaging via the HRSC and atmospheric profiling through the Planetary Fourier Spectrometer (PFS). These instruments detected cloud trains and lee wave structures near Arsia Mons and other volcanic complexes [9]. Most recently, the Emirates Mars Mission (EMM), launched in 2020 and operational since 2021, has provided a new level of atmospheric coverage through its payload of instruments, including the Emirates eXploration Imager (EXI), the Emirates Mars Infrared Spectrometer (EMIRS), and the Emirates Mars Ultraviolet Spectrometer (EMUS). EMM has provided unprecedented diurnal and seasonal monitoring of atmospheric phenomena, enabling detailed observation of lee wave clouds across varying latitudes and times of day [10]. Together, these missions showcase significant advances in observing Mars and highlight how EMM is setting a new standard in tracking and understanding lee wave clouds.

## 3. Occurrences and Locations

Cloud observations on Mars provide a wealth of information regarding local meteorology, weather systems, and seasonal climatic variations. The earliest observations of clouds were detected by Mariner 9 and Viking orbiter missions back in 1979, in which cloud occurrences with seasons were recorded and observed on specific locations on Mars since then [5]. It has been observed that lee wave clouds usually occur in mid-to-high latitudes on Mars and are rarely observed in the equator.

As for the season, these occurrences usually occur during the spring and fall-winter time [1]. A complete assessment of occurrences per season has been conducted using the Mariner 9 and Viking orbiter images for MY 13 and 14. Observations show that during $L_S = 0°-90°$, lee wave clouds were seen to be abundant at mid-to-high southern latitudes and during $L_S = 90°-180°$ in the southern mid-latitude and northern hemisphere [5]. Similar observations were noted using the Mars Color Camera (MCC) images during years MY 33 and 34 in which the amount of lee wave clouds could be seen near $L_S=90°-110°$ reaching its peak near $L_S=100°$ with it decreasing to a minimum near $L_S=140°$ [2]. In contrast, the Mars Global Surveyor (MGS), using MOC images, documents that there is little occurrence of lee wave clouds before $L_S=170°$, captured during MY 26 [11]. For $L_S=180°-270°$, there were only a few observations of lee wave clouds during this season observed by Viking and Mariner 9. While a notable lee wave cloud occurrence was captured by MGS, generated by Milankovic crater near the Northern polar region on $L_S=234°$ [11]. As the season progresses towards winter $L_S=270°-360°$, it has been observed by Mariner 9 an abundance of lee wave clouds in the north, which might also be triggered by a global dust storm event that occurred during that time [5]. During this season, it has been observed that Mars exhibits significant lee wave activity in the mid-latitudes during fall and winter, as the wind shear is very high, with wind speeds increasing dramatically with height due to the winter mid-latitude jet stream. This strong fluctuation of wind with height is only present during these seasons [11]. Table 1 highlights a summary of these findings in table format.

Table 1. The Table showcases the availability of lee Wave clouds based on the missions that captured observations for MY 13, 14, 26, 33, 34, and 36, which is the focus of this study. These findings were reported by French, 1981 [5], Kalita et al., 2022 [4] and Wood et al. 2003 [11].

| | Ls (Solar Longtitude) | Ls = 0°-90° | Ls = 90°-180° | Ls = 180°-270° | Ls = 270°-360° | Mission |
|---|---|---|---|---|---|---|
| Mars Year (MY) | MY 13 | mid-to-high Southern Latitudes | Southern mid-latidues and Northern hemisphere | ------------------ | mid-latitudes | Mariner 9 and Viking |
| | MY 14 | ------------------ | ------------------ | ------------------ | mid-latitudes | Mariner 9 and Viking |
| | MY 26 | ------------------ | ------------------ | Milankovic craternear Northern Polar region | ------------------ | Mars Global Surveyor (MGS) |
| | MY 33 | mid-to-high Southern Latitudes | Southern mid-latidues and Northern hemisphere | ------------------ | ------------------ | Mars Orbiter Mission (MOM) |
| | MY 34 | mid-to-high Southern Latitudes | Southern mid-latidues and Northern hemisphere | ------------------ | ------------------ | Mars Orbiter Mission (MOM) |
| | MY 36 | Southern mid-latidues and Northern hemisphere | ------------------ | Mid-latitudes Northern Part | Mid-latitudes Northern Part | Emirates Mars Missions |

## 4. Methodology

Initially, we aim to filter the EXI L2A 635nm band data that contained clouds with lee wave pattern from the full MY 36. Filtering the data was attempted through various approaches for automation, ranging from simple classification techniques for cloud detection to Machine Learning with specific cloud patterns that lee wave clouds exhibit. However, the results were not as reliable as they needed to be. Out of 3706 EXI L2A 635nm band images processed, we have identified 591 images to contain potential lee wave clouds. Through manual verification, we have concluded that the tool is more suitable for general cloud detection rather than specifically for lee wave clouds, as it did not successfully capture the unique cloud pattern and required adjustments. We have briefly tested GPT-4, as using AI could potentially resolve these issues; however, due to time constraints, we resorted to the manual approach. Utilizing AI capabilities for image classifications using deep learning methods is an extended topic for future work. Recent advances in deep learning have demonstrated promising capabilities in the automated identification of orographic cloud phenomena. Coney et al. (2024) [12] developed a convolutional neural network model trained on hand-labeled vertical velocity fields to detect and characterize trapped Lee waves with high precision. Their method effectively segments lee wave structures and enhances the spatial understanding of their formation over complex terrain. While their approach focuses on Earth-based Numerical Weather Prediction (NWP) data, such methodologies offer strong potential for adaptation to Martian cloud detection studies. In future work, we can consider the application of similar deep learning frameworks, particularly for the automated recognition of lee wave cloud trains in imagery from missions such as EMM.

## 5. Results

Along the full MY 36, lee waves clouds were found in total of 50 EXI image observations. Our findings are unique in terms of capturing observations at different times of day which is crucial in the overall understanding of how lee waves clouds behave during the day and the different timing they can develop. The images were identified initially using machine learning mechanisms. However, with many challenges to perfect the algorithm to detection of lee wave clouds (as highlighted in the Methodology section of this paper), the images were then filtered out manually. Based on the observations, we found that during $L_S=0°-90°$, lee wave clouds were found in the mid-latitude region between 22° latitude and ~ -20°. As we went further into the season, the lee wave clouds were observed in the Northern part initially at the start of the season. As the season moves towards $L_S=0°-90°$, lee waves were observed towards the southern part of the hemisphere. As for the timing, lee wave clouds were found in the afternoon time (between >12:00 - 19:00) and a few observations that identify lee waves clouds during the evening time (>19:00 – 00:00). During $L_S$=90°-180°, there were no detections of lee wave clouds, which is consistent to the other mission findings that lee wave clouds occurrences decrease during this season. For $L_S$=180°-270°, only one observation was found in mid-latitude Northern part around the Arabia Terra during the evening time (>19:00 – 00:00). Viking and Mariner 9 observation is similar with little activity being observed during MY 26. As the season progresses to $L_S$=270°-360°, an abundance of lee wave clouds can be seen during the morning time (>00:00 – 12:00) and afternoon time (between >12:00 – 19:00) in the mid-latitude Northern hemisphere. These results are very consistent with the other missions that observed lee wave clouds dating from MY 13 to MY34.

## 6. Discussion

The seasonal and spatial pattern of lee wave clouds on Mars is consistent with previous missions' data. The lee wave clouds appeared mostly at mid-latitudes from 22°N to ~20°S during the early Martian season (Ls=0°–90°) with cloud activity beginning in the northern hemisphere and progressively migrating southward with season. Chronologically, the clouds were observed more frequently in the afternoon, although some were also observed in the evening. Importantly, no lee wave clouds were identified during Ls = 90°–180°, consistent with previous observations noting a reduction in activity during this time frame. For Ls =180°–270°, no more than a single case of lee wave cloud formation is detected in the northern mid-latitudes at noon. In contrast, as the season progressed to Ls = 270°–360°, lee wave cloud activity picked up again, concentrated in the northern mid-latitudes during both the morning and afternoon hours. Figure I shows the number of observations collected during the time of day per season for MY 36, as captured by the EMM mission. Based on the results, lee wave clouds have been observed to have the most activity during the afternoon time between 12:00 – 19:00. The most activity was found to be during Ls = 270°–360°.

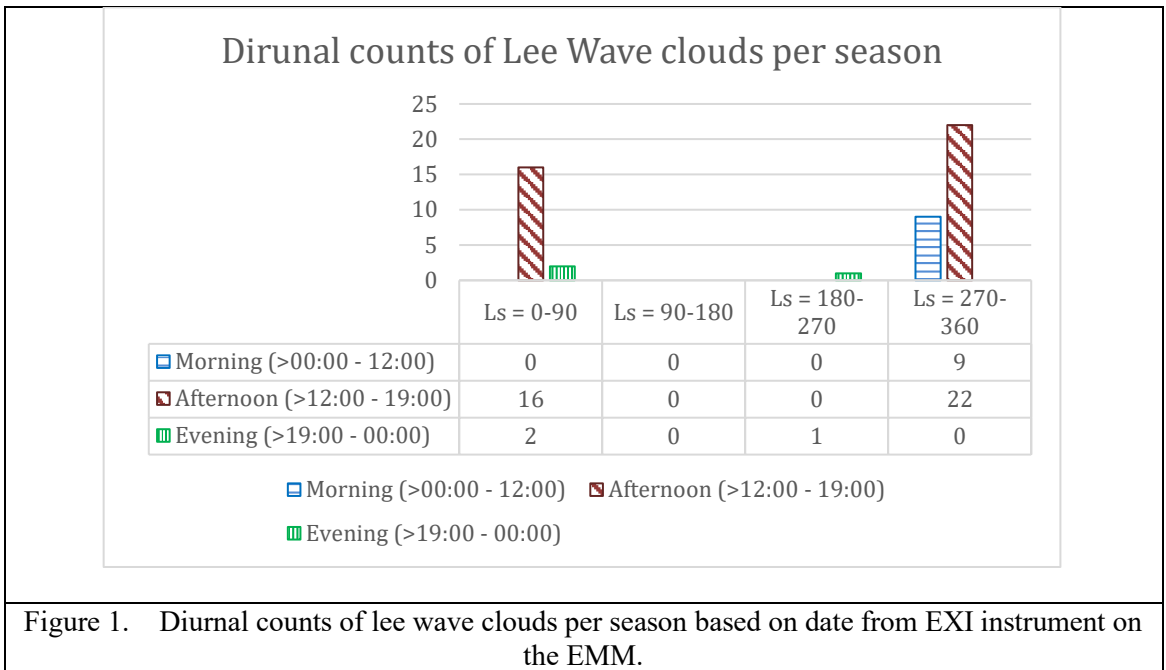

Figure 1.  Diurnal counts of lee wave clouds per season based on date from EXI instrument on the EMM.

## 7. Conclusion

Assessing the identification of lee wave clouds during MY 36 using the EMM dataset provided valuable context for understanding the spatial and seasonal distribution of these phenomena. By analyzing red band imagery from EXI, we confirmed 50 distinct lee wave cloud occurrences, with peak activity observed during the afternoon hours in the late southern spring to early summer (Ls =270°–360°). Comparisons with previous mission data validated our detection approach and underscored the significance of diurnal evolution in lee wave behavior. While this study focused on capturing the global characteristics of lee-wave cloud activity across the whole study period, a logical next step is to refine our analysis to focus on specific regions and events. In particular, the work by Gebhardt et al. (2023) [13], which presents sub-hourly EXI observations of a rapidly evolving dust storm and associated lee wave clouds near Lyot Crater, demonstrates the power of high-frequency, location-focused monitoring. Their findings reveal how morning gravity-wave patterns evolve into afternoon convective dust activity, which emphasizes the dynamic coupling between topography and atmospheric processes. Our future efforts will focus on prioritizing site-specific case studies aimed at predicting lee-wave formation and diurnal transformation, thereby deepening our understanding of Martian mesoscale meteorology.

## 8. Acknowledgments

The authors would like to thank Professor Micheal Wolff for the scientific idea of the research and MBRSC as well as UAESA for publicly providing access to the UAE EMM data.